\documentclass[aps,prd,preprintnumbers,superscriptaddress,nofootinbib,onecolumn,notitlepage]{revtex4-1}
\usepackage{amsmath,amssymb}
\usepackage[colorlinks=true,linkcolor=magenta,citecolor=magenta]{hyperref}

\begin{document}
\title{Stable cosmological solutions in degenerate theory of gravity}

\author{Ryotaro Kase}
\affiliation{Department of Physics, Faculty of Science, Tokyo University of Science, 1-3, Kagurazaka, Shinjuku-ku, Tokyo 162-8601, Japan}
\author{Rampei Kimura}
\affiliation{Department of Physics, Tokyo Institute of Technology, 2-12-1 Ookayama, Meguro-ku, Tokyo 152-8551, Japan}
\author{Atsushi Naruko}
\affiliation{Frontier Research Institute for Interdisciplinary Sciences \& 
Department of Physics, Tohoku University \\
Aramaki aza Aoba 6-3, Aoba-ku, Sendai 980-8578, Japan}
\author{Daisuke Yoshida}
\affiliation{Department of Physics, McGill University, Montr\'eal, Qu\'{e}bec, H3A 2T8, Canada}

\date{\today}

\preprint{TU-1058}

\begin{abstract}
We study cosmological applications of extended vector-tensor theories,
whose Lagrangians contain up to two derivatives with respect to metric and vector field.
We derive background equations 
under the assumption of homogeneous and isotropic universe
and study the nature of cosmological perturbations on top of the background.
We found an example of healthy cosmological solutions in non-trivial degenerate theory of gravity for the first time,
where those perturbations do not suffer from any instabilities, that is, 
ghost and gradient instabilities. 
\end{abstract}

\maketitle

\section{Introduction}
Although general relativity may be the most successful theory for describing gravitational interactions
and consistent with observations and experiments at solar-system scales with high accuracy \cite{Will:2014kxa},
it is not sure whether general relativity can be applied to larger scales, {\it e.g.}, cosmological scales.
In fact, motivated from the fact that the universe was/is undergoing accelerated phases of expansion 
at early \cite{Brout:1977ix,Starobinsky:1979ty,Sato:1980yn,Guth:1980zm,Linde:1981mu} and recent epoch \cite{Riess:aa,Perlmutter:aa}, 
alternative gravitational theories beyond general relativity have attracted considerable attention
for the last several decades.

A simple way to modify general relativity will be to introduce a scalar degree of freedom,
which could
non-minimally couple with gravitational or matter field, 
and it then mediates the fifth force. 
Here, we have to be careful because such a modification could induce the change of gravitational law at all scales.
Indeed, the first example of modified gravity, Brans-Dicke theory \cite{PhysRev.124.925}, 
is tightly constrained from the solar-system tests \cite{Will:2014kxa}. 
Under the constraint, it is difficult to find difference from general relativity. 
However, there are viable models that can evade these constraints,
that is, modification of gravity only takes place at cosmological scales (see for reviews {\it e.g.} \cite{Clifton:2011aa,Tsujikawa:2011aa,Koyama:2015vza}).
In such models, the law of gravity reduces to the one in general relativity
through screening effects of the fifth forth at small scales.
In one of the famous screening effects called Vainshtein mechanism \cite{Vainshtein:1972aa,Babichev:2013usa},
the non-linearities of scalar self-interactions containing higher derivatives become important at small scales,
and as a result the scalar field effectively weakly couples with gravity. 

On the other hand, once equations of motion have higher-order time
derivatives (more than third derivatives),
the system, in general, contains extra degrees of freedom known as Ostrogradsky's ghosts \cite{Ostrogradsky1850}. 
Therefore, one should design a theory with care when a Lagrangian contains higher derivatives
as in the case of a galileon field \cite{2009PhRvD..79h4003D,Nicolis:2008aa}. 
In galileon theories, although the Lagrangian contains second derivatives, 
which cannot be removed by integrating by parts, 
the equations of motion for gravity and scalar field successfully
remain second-order differential equations. 
The galileon field can be further extended, and the most general scalar-tensor theory, whose equations of motion are at most second-order differential equations,
is now known as Horndeski theory \cite{Horndeski:1974aa} (or Generalized galileon \cite{Deffayet:2011aa,2011PThPh.126..511K}, see for review e.g. \cite{Deffayet:2013lga}). 

Intriguingly, it has been recently argued that one can construct a theory without Ostrogradsky's ghost 
even if Euler-Lagrange equations contains higher derivatives, as long as the theory has an appropriate number of constraints (see \cite{Motohashi:2016ftl,Motohashi:2017eya} for bosons and \cite{Kimura:2017gcy} for fermions).
The first example has been found in the context of the curved space-time extension of galileon theory 
called beyond Horndeski or GLPV theory \cite{2015PhRvL.114u1101G}. 
The generalization of degenerate higher-order scalar-tensor theories has been recently investigated 
in the case of the Lagrangian that depends on second order derivatives 
quadratically \cite{2016JCAP...02..034L,2015arXiv151206820L,2016arXiv160208398B,2016JCAP...04..044C} 
and cubically \cite{BenAchour:2016fzp}. Furthermore, 
the stability analysis of the quadratic theory on a cosmological background has been investigated in \cite{2016arXiv160408638D}.
Unfortunately, the tensor modes are absent in most cases  
except for the cases that can be mapped from Horndeski theories 
via conformal and disformal metric transformations. 
Even for the remaining class, there are instabilities in either scalar 
or tensor perturbation.
 
Recently, three of the authors in the present paper formulated 
yet another degenerate theories of gravity coupled with a massive/massless vector field \cite{Kimura:2016rzw}. 
This extended vector-tensor theories contain up to two derivatives with respect to 
metric and vector field, and the degeneracy of the kinetic matrix ensures that 
the dynamical degrees of freedom in the theory are at most $2+3$ for gravity and vector sectors. 
These theories include the generalized Proca theories 
\cite{Heisenberg:2014rta,Tasinato:2014eka,Tasinato:2014mia,Jimenez:2016isa} 
and those extensions (beyond generalized Proca theories) 
\cite{Heisenberg:2016eld}.
Although in extended vector-tensor theories 
one component of vector field can be removed by a constraint, 
one need to check whether the theory is healthy or not, {\it i.e.,} all modes 
can propagate
on a background of interest without suffering any instabilities.
In this paper, we explore this stability issues of perturbations in the extended vector-tensor theories
around a homogeneous and isotropic background.
Throughout this paper, 
we use units in which the speed of light and the reduced Planck mass are unity.

\section{Theory}
The most general action of vector-tensor theory that contains up to two derivatives for the metric $g_{\mu\nu}$ and the vector field $A_\mu$ respecting 4-dimensional general covariance, is given by 
\begin{align}
 S &= \frac{1}{2}\int d^4 x \sqrt{-g} \Bigl[ f(Y) R + C^{\mu\nu\rho\sigma}\nabla_{\mu}A_{\nu}\nabla_{\rho}A_{\sigma} + G_3(Y)\nabla_{\mu}A^{\mu} + G_2(Y)   \Bigr] 
 + \int d^4 x \sqrt{-g} {\cal L}^m \,,
\end{align}
where $R$ is the Ricci scalar, $\nabla_\mu$ is a covariant derivative with respect to the metric $g_{\mu\nu}$,
${\cal L}^{m}$ is the Lagrangian of matter field, and
$C^{\mu\nu\rho\sigma}$ is defined by
\begin{align}
 C^{\mu\nu\rho\sigma} =&
 	\alpha_1(Y) g^{\mu(\rho}g^{\sigma)\nu} + \alpha_2(Y) g^{\mu\nu}g^{\rho\sigma}
	 + \frac{1}{2}\alpha_3(Y) (A^\mu A^\nu g^{\rho\sigma} + A^\rho A^\sigma g^{\mu\nu}) \notag\\
	& \qquad
	  + \frac{1}{2} \alpha_4(Y) (A^\mu A^{(\rho} g^{\sigma)\nu} + A^\nu A^{(\rho} g^{\sigma)\mu}) 
	  + \alpha_5(Y) A^{\mu} A^{\nu} A^{\rho} A^{\sigma} + \alpha_6(Y) g^{\mu[\rho}g^{\sigma]\nu}\notag\\
	& \qquad
	+ \frac{1}{2}\alpha_7(Y) (A^\mu A^{[\rho} g^{\sigma]\nu} - A^\nu A^{[\rho} g^{\sigma]\mu})
	+ \frac{1}{4}\alpha_8(Y) (A^\mu A^\rho g^{\nu\sigma} - A^{\nu}A^{\sigma} g^{\mu\rho})
	+\frac{1}{2}\alpha_9(Y) \epsilon^{\mu\nu\rho\sigma} \,.
\end{align}
Here $f$, $\alpha_i$, and $G_i$ are arbitrary functions of Proca mass term $Y = A_{\mu}A^{\mu}$.
The first five terms with $\alpha_{1,2,3,4, 5}$ represent symmetric parts of the tensor $C^{\mu\nu\rho\sigma}$, namely the tensor is invariant under the change of indices such as $\mu \leftrightarrow \nu$ and $\rho \leftrightarrow \sigma$ while the remaining terms are antisymmetric.
Generally, this theory contains $2+4$ degrees of freedom unless degeneracy conditions are satisfied.
According to \cite{Kimura:2016rzw}, one can find the theories, whose degrees of freedom are at most five, 
by imposing degeneracy conditions in a general curved background. 

In the present paper, we mainly focus on the case B5,
in which all perturbations are free of ghost and gradient instabilities,
as we will see. 
The arbitrary functions in the case $B5$ are given by
\begin{eqnarray}
\alpha_1&=&\frac{
	-8(2\alpha_2+Y\alpha_3)+Y(-4-4Y\alpha_2+Y^2\alpha_3)\alpha_8}{2Y^2\alpha_8} ,\notag
\\
\alpha_4&=&{4(1+Y\alpha_2) \over Y^2}-\alpha_3+{8(2\alpha_2+Y\alpha_3) \over Y^3 \alpha_8}+\alpha_8-{Y^2\alpha_8^2 \over 8} ,\notag
\\
\alpha_5&=&{-2+Y^2 \alpha_3 \over Y^3}-{4(2\alpha_2+Y\alpha_3) \over Y^4 \alpha_8}-{\alpha_8 \over Y}+{Y\alpha_8^2 \over 8}
+{12(2\alpha_2+Y\alpha_3) \over Y^2(-8+Y^2\alpha_8)} \, .
\label{B5}
\end{eqnarray}
It should be noted that $\alpha_8$ is non-vanishing in this branch as well as 
$Y^2 \alpha_8 \neq 8,16$ and $Y \alpha_1 \neq 1$. 
Due to these conditions, we cannot smoothly switch off the antisymmetric parts. 
Hence we cannot reproduce any degenerate scalar-tensor theory under the direct replacement of $A_\mu$ by $\nabla_\mu \phi$, in other words, 
there is no direct counterpart of this branch in any known theory of scalar field.
Here, $\{\alpha_2, \alpha_3, \alpha_7, \alpha_8\}$
are free arbitrary functions, and we have set $f=1$ and $\beta=-2\alpha_6-\alpha_7 Y=0$ 
without loss of generality due to the properties of
the conformal and disformal metric transformations\footnote{
Note that this is only true when matter fields are absent.}. 
The degeneracy combinations of other cases can be found in \cite{Kimura:2016rzw}. \\

\section{Background and Perturbations}
\subsection{Background}
Let us consider a flat Friedmann-Lema\^itre-Robertson-Walker (FLRW) metric :
\begin{align}
 ds^2 
 = - N^2(t) \, dt^2 + a^2(t) \delta_{ij} \, dx^{i}dx^{j}
 = - d \tau^2 + a^2(\tau) \delta_{ij}dx^{i}dx^{j} \,,
\end{align}
where the proper time $\tau$ can be related to the cosmic time by $d \tau = N(t) d t$,
and homogeneous and isotropic vector field 
will be described by
$\bar{A}_{\mu}dx^{\mu} = A_*(\tau) d\tau = A_* (t) N(t) dt$.
Then, the background action in the case $B5$ is given by
\begin{align}
\bar{S} &= 
\int d t \,d^3 x\ N a^3
 \left[\frac{1}{2}  \left\{
 -{\lambda \over N^2} \left( 
 \frac{1}{a}
 {d a \over dt}
 +\omega 
 {d A_* \over dt}
 \right)^2  +G_2  
 -{G_3 \over N a^3} {d\over dt} \biggl(a^3 A_* \biggr)\right\}
+ {\cal \bar L}^m
 \right] \,,
 \label{backN}
\end{align}
where we have defined $\lambda$ and $\omega$ as 
\begin{align}
 \lambda(-A_{*}^2) &= \frac{3( \alpha_3 A_*^2 - 2 \alpha_2 )( \alpha_8 A_*^4 - 8)}{ 2  \alpha_8 A_*^2 }, \\
 \omega(-A_*^2) 
 &= \frac{\alpha_8 A_*^3}{\alpha_8 A_*^4  - 8}.
\end{align}
The equations of motion of the lapse function and vector field can be 
derived by varying the action with respect to $N$ and $A_*$,
\begin{align}
0  
&= \lambda (H + \omega \dot{A}_{*})^2 +  G_2 - 2 \rho \,,
\label{Friedmann}\\
 0 
&= 3 G_3' A_*^2 H - G_2' A_* 
+ A_* \biggl[ \lambda ' H^2- \omega  \left(\omega  \lambda '+2 \lambda  \omega '\right)\dot{A}_*^2\biggr]
+ \lambda  \omega \biggl[\dot{H}+\omega  \ddot{A}_*+3 H \left(H+\omega\dot{A}_*  \right)\biggr] \,,
\label{EOMA0}
\end{align}
where a dot represents a derivative with respect to the proper time, 
$H = \dot{a}(\tau)/a(\tau)$, and a prime denotes a derivative with respect to $\bar{Y}:=-A_*^2$.
Here, we have introduced $\rho$, the energy density for the matter.  
The energy conservation equation for the matter field is given by 
\begin{eqnarray}
\dot \rho + 3H (\rho +P) =0\,,
\label{EMC}
\end{eqnarray}
where the energy density $\rho$ and the pressure $P$ for the matter are defined by 
\begin{eqnarray}
\rho := -{1 \over  a^3}  {\delta (N a^3\bar{{\cal L}}^m )\over \delta N} \,,  \qquad
P :=  {1 \over  3 N a^2 } {\delta (N a^3\bar{{\cal L}}^m ) \over \delta a} \,.  
\end{eqnarray} 

Taking a proper time derivative of (\ref{Friedmann}), we obtain
\begin{eqnarray}
 \left(H+\omega \dot{A}_*  \right) \biggl[\lambda  \left(\dot{H}+\omega  \ddot{A}_*-2 \omega ' A_* \dot{A}_*^2 \right)
- \lambda '  A_* \dot{A}_*\left(H+\dot{A}_* \omega \right)\biggr]-  G_2' A_* \dot{A}_* -\dot{\rho}=0\,.
\label{EOMNt}
\end{eqnarray}
Combining (\ref{EOMA0}) and (\ref{EOMNt}), 
we can eliminate both $\ddot{A}_*$ and $\dot{H}$ and obtain 
\begin{eqnarray}
H\biggl[ 
3 \omega (\rho +P)
 + G_2'  A_{*}  -3G_3' A_{*}^2 \left(H+\omega\dot{A}_*  \right) 
-\left(3 \lambda  \omega +A_{*} \lambda'\right)\left(H+\omega\dot{A}_*  \right)^2 \biggr]=0\,.
\label{EOMA0n}
\end{eqnarray}
As can be seen from (\ref{Friedmann}) and (\ref{EOMA0n}), the background equations are reduced to the first-order differential equations of $A_{*}$ and $H$. 
One can also derive the equation of motion for $a$ from (\ref{backN}), however,
it is enough to use the independent equations (\ref{Friedmann}), (\ref{EMC}), and (\ref{EOMA0n}).
Assuming $H\neq 0$, we have the following two cosmological solutions
depending on whether $\lambda$ is zero or non-zero.
Here, $\lambda$ vanishes when and only when $2\alpha_2+Y\alpha_3 =0$
since $Y^2 \alpha_8 \neq 8$ in the case of $B5$.

\vspace{4mm}
{\bf Solution I} ($\lambda=0$) : 
From (\ref{Friedmann}), one immediately finds
\begin{eqnarray}
G_2(-A_*^2) = 2 \rho\,, 
\label{sol1.1matter}
\end{eqnarray}
which implies that $A_*$ is a function of $\rho$. 
Solving Eq.~(\ref{EOMA0}) in terms of $H$, we obtain
\begin{eqnarray}
&&H = \frac{G'_2(-A_{*}^2)}{3A_* G'_3(-A_{*}^2)}\,.
\label{sol1.2}
\end{eqnarray}
Since $A_{*}$ is a function of energy density $\rho$, Eq. \eqref{sol1.2} can be regarded as a modified Friedmann equation, which 
determines the Hubble function as a function of energy density of matter field.
For simplicity, we will focus on the case where the matter component is the cosmological constant, $\rho = - P = \Lambda$.
Then Eq. \eqref{sol1.1matter} simply gives that $A_*$ is constant and hence
 $H$ is also constant, which leads to de Sitter expansion. 

Note that if one uses the following power law functions,
\begin{align}
G_2(Y) = g_2 (- Y)^{(2n + 1)}\,,\qquad G_3(Y) = g_3 (-Y)^{n}\,,
\label{arbitrary1}
\end{align}
Eq. \eqref{sol1.2}
takes the familiar form of the Friedmann equation,
\begin{align}
 H^2 = \frac{\rho}{3 M_{\rm eff}^2},
\end{align} 
where the effective Planck mass $M_{\rm eff}$ is given by
\begin{align}
 M_{\rm eff}^2 =  \frac{3 g_3^2 n^2}{2 g_2 (1 + 2 n)^2}  \,.
\label{Meff}
\end{align}

\vspace{4mm}
{\bf Solution II} ($\lambda\neq 0$) : 
We first solve Eq.~(\ref{Friedmann}) for $H+\omega\dot{A}_*$
and substitute the solution into Eq.~(\ref{EOMA0n}). 
Then, we obtain the equation, which determines $A_*$ algebraically
\begin{eqnarray}
&& 
3 \omega (\rho +P)
+ G_2' A_* -3 G_3' A_*^2  \sqrt{\frac{2 \rho -G_2}{\lambda }}  
- (3\lambda \omega +A_* \lambda')\frac{2 \rho -G_2}{\lambda}=0\,.
\label{sol2.1}
\end{eqnarray}
Since $A_*$ can be solved as a function of $\rho$ and $P$, Eq.~(\ref{Friedmann}) then determines $H$ as a function of $\rho$ and $P$
\begin{eqnarray}
H=\sqrt{\frac{2\rho-G_2}{\lambda}} - \omega \dot{A}_{*}\,.
\label{sol2.2}
\end{eqnarray}

\subsection{Perturbations}
In the present paper, we would like to demonstrate the existence of perturbatively healthy 
cosmological solutions in degenerate theory of gravity for the first time.
The analysis in this paper is restricted to de Sitter background driven by the cosmological constant in the solution I of the case $B5$  for the sake of simplicity. 
We have also performed the full analysis of perturbations even in the presence of 
matter that can be reported in the subsequent paper.
Hereinafter the lapse function is set to be unity in which the proper time exactly coincides with the cosmic time and hence a dot can be regarded as a derivative with respect to the cosmic time.
Also, the perturbations for the metric and the vector field around a background are defined as 
	\begin{align}
	& g_{\mu\nu} = \bar{g}_{\mu\nu} + \delta g_{\mu\nu}\,, \\
	& A_{\mu} = \bar{A}_{\mu} + \delta A_{\mu}\,. 
	\end{align}

\vspace{4mm}
{\bf Tensor perturbations }
The tensor perturbations in the metric and vector field are defined by
\begin{align}
 \delta g_{\mu\nu}dx^{\mu}dx^{\nu} &= h_{ij} dx^{i}dx^{j}\,,\\
 \delta A_{\mu} &= 0\,,
\end{align}
where $h_{ij}$ is traceless and transverse, 
which satisfies $h_{ii} = \partial_{i}h_{ij} = 0$.
The Fourier components of tensor perturbations are given by,
\begin{align}
 h_{ij}(t, \mathbf{x}) =
 \int \frac{d^3 k}{(2\pi)^3} \Bigl[h_{+}(t,\mathbf{k}) e^{+}_{ij}(\mathbf{k})  
+ h_{\times}(t,\mathbf{k}) e^{\times}_{ij}(\mathbf{k})
 \Bigr] \mathrm{e}^{i \mathbf{k} \cdot \mathbf{x}}\,,
\end{align}
where $e_{ij}^{+}(\mathbf{k})$ and $e_{ij}^{\times}(\mathbf{k})$ are orthonormal and traceless bases of 
the tensor field space orthogonal to $\mathbf{k}$. 
Here, $\mathbf{x}=(x^1,x^2,x^3)$ and $\mathbf{k}=(k_1,k_2,k_3)$.
The second order action can be straightforwardly obtained by 
expanding the action up to quadratic order,
\begin{align}
 S^{(2)}_{\rm tensor} = \int dt \frac{d^3 k}{(2\pi)^3} \sum_{I={+,\times }}
 a^3 \, {\cal K}_T \left( \dot{h}_{I}{}^2 - c^2_T \frac{k^2}{a^2}  h^2_{I}    \right)\,,
\end{align}
where coefficients are given by
\begin{align}
{\cal K}_T = \frac{1}{4} ( 1 + A_*^2 \alpha_1), \quad
c^2_T =\frac{1}{1 + A_*^2 \alpha_1}\,.
\label{tensorcon}
\end{align}
Here we have used short hand notation like $h_{I}{}^2 := h_{I}(t, \mathbf{k}) h_{I} (t, - \mathbf{k}) $.

Before ending this section, it will be interesting to see that 
the speed of gravitational waves can be set unity in this model.
From the recent observation of gravitational-wave event GW170817
\cite{TheLIGOScientific:2017qsa} with the promising counterpart 
gamma-ray burst GRB 170817A \cite{Goldstein:2017mmi}, 
the bound on the sound speed of tensor mode can be obtained as
$-3 \times 10^{-15}\leq c_T - 1 \leq 7\times 10^{-16}$ 
\cite{Monitor:2017mdv}. 
By setting $\alpha_1 = 0$, the arbitrary functions $\alpha_2$
	and $\alpha_3$ are fixed as
\begin{eqnarray}
\alpha_2=-\frac{2}{3Y}\,,\qquad
\alpha_3 =\frac{4}{3Y^2} \,,
\label{arbitrary0}
\end{eqnarray}
for $\lambda =0$. 

\vspace{4mm}
{\bf Vector perturbations }
We use the spatially flat gauge, 
where the spatial components of metric perturbations vanish by the gauge condition, i.e.,
\begin{eqnarray}
\delta g_{\mu\nu}dx^{\mu}dx^{\nu} &=& 2 a S_{i} dt dx^{i}\,, 
\end{eqnarray}
and we define the vector field perturbation as 
\begin{eqnarray}
\delta A_{\mu}dx^{\mu} &=&  \delta A_{i}dx^{i} \,.
\end{eqnarray}
Here $S_i$ and $\delta A_{i}$ satisfy the conditions $\partial_{i} S_{i} = \partial_{i}\delta A_{i} = 0$. 
The Fourier modes of vector perturbations are defined by
\begin{align}
 V_i(t, \mathbf{x}) = \int \frac{d^3 k}{(2 \pi)^3} \sum_{I=1,2} V_I(t, \mathbf{k}) e_i^{I}(\mathbf{k}) \mathbf{e}^{i \mathbf{k}\cdot \mathbf{x}}\,, 
\end{align}
where $V_i$ ($V_I$) stands for $S_i$ ($S_I$) or $\delta A_i$ ($\delta A_I$). $e_{i}^{1}(\mathbf{k})$ and $e_{i}^{2}(\mathbf{k})$ are orthogonal bases of the vector field space which orthogonal to $\mathbf{k}$.
Thanks to the general covariance of the original action and the fact that the theory does not involve any higher derivative of metric/vector field, one can integrate out $S_I$, and the resultant action for $\delta A_I$ is found to be 
\begin{align}
 S^{(2)}_{\rm vector} = \int dt \frac{d^3 k}{(2 \pi)^3} \sum_{I = 1,2 } {
 a^3 \,
{\cal K}_V\left(
 \dot{\widetilde{\delta A}}_{I}^2 - {\cal M}_V \widetilde{\delta A}_{I}^2
 \right)
 } \,,\label{SV}
\end{align}
where $\widetilde{\delta A}_I = \delta A_I/a$. The coefficients ${\cal K}_V$ and $ {\cal M}_V$ in the limit of large $k/a$ are given by
\begin{align}
 \mathcal{K}_V&= {(\alpha_4+\alpha_8)A_*^2-2\alpha_1\over 4} \,, 
 \label{KV} \\
 {\cal M}_V &= { 
c_V{}^2 \, \frac{k^2}{a^2} \Biggl[ 1 + {\cal O} \Bigl( (a/k)^2 \Bigr) \Biggr] \,, 
 }
\end{align}
with
\begin{align}
c_V{}^2&= -\frac{\alpha_1(2+\alpha_7 A_*^4) + \alpha_7 A_*^2
}{(1+\alpha_1 A_*^2) \bigl[(\alpha_4+\alpha_8) A_*^2-2\alpha_1 \bigr]}\,.
\label{CV}
\end{align}

\vspace{4mm}
{\bf Scalar Perturbations }
Now we define the metric and vector field perturbations in the spatially flat gauge as
\begin{align}
\delta g_{\mu\nu}dx^{\mu}dx^{\nu}  &= -  2 \phi dt^2 + 2 a \partial_{i} B dx^{i}\,,\\
 \delta A_{\mu} dx^{\mu} &= (\delta A_{*} + A_{*} \phi) dt  + \partial_{i} \delta A_{L} dx^{i}\,. 
\end{align}
Similar to the tensor and vector perturbations, 
we then define Fourier mode of scalar perturbations by
\begin{align}
 \Phi^{I}(t, \mathbf{x}) = \int \frac{d k^3}{(2 \pi)^3}  \Phi^{I}(t,\mathbf{k}) \mathrm{e}^{i \mathbf{k} \cdot \mathbf{x}}\,,
\end{align}
where $\Phi^{I}$ represent scalar perturbations, $\{\phi, B, \delta A_*, \delta A_L\}$.
Hereafter, to simplify the analysis, we shall set the sound speed of tensor mode to be unity, 
	$c_T=1$ demanding $\alpha_1 = 0$ or equivalently $\alpha_2=-2/(3Y)$ and $\alpha_3 = 4/(3Y^2)$ as found in (\ref{arbitrary0}).
With the use of the background equations, 
one can confirm that the variables $\phi, B$, and $\delta A_*$
carry no derivative in the quadratic action, and we then obtain 
three constraints for them. After integrating out these variables\footnote{
	One should note that there exist special cases in which
	some of the coefficients of $\phi$, $B$, and $\delta A_*$ in constraint equations 
	vanishes identically due to special choices of arbitrary functions or 
	background solutions. 
	In such cases, one needs to carefully solve the constraint equations,
	and different results might be obtained. 
	However, we disregard such special situations in the present analysis.
}, 
we finally obtain the reduced action for $\delta A_L$, 
which can be written as,
\begin{align}
	S_{\rm scalar}^{(2)} = \int d t \frac{d^3 k}{(2 \pi)^3} 
	{a^3 \, {\cal K}_S \left(   \dot{\widetilde{\delta A}}_L{}^2 - {\cal M}_S  \, \widetilde{\delta A}_L{}^2 \right) \,,}
\end{align}
where $\widetilde{\delta A}_L = (a/k)^2 \delta A_L$. The coefficients in the limit of large $k/a$ are given by
\begin{align}
	{\cal K}_S = \frac{G'^2_2 A_* (\omega A_* -2)}{2\omega} + {\cal O} \Bigl( (a/k)^2 \Bigr)\,,
\label{KS}
\end{align} 
and
\begin{align}
	{\cal M}_S = c_S^2\frac{k^2}{a^2} \Biggl[ 1 + {\cal O} \Bigl( (a/k)^2 \Bigr) \Biggr] \,,
\end{align}
with the sound speed of the scalar mode, 
\begin{align}
c_S^2 &= \frac{1}{9\omega A_*^5 ( \omega A_*-2 )
	G'^3_3}
\biggl[ 3G'^3_3 A_*^4 - 4 G''_2 G'_3 A_*^2 + 4 G'_2 G''_3 A_*^2 
-2G'_2 G'_3 (1+\omega A_*)
\biggr]\,.
\label{CS}
\end{align}

\section{Stabilities}
The necessary conditions for avoiding ghost and gradient instabilities are given by 
\begin{eqnarray}
&&{\cal K}_T, \,
{\cal K}_V, \,
{\cal K}_S, \,
> 0\,,  \quad 
c_T^2, \,
c_V^2, \,
c_S^2, \, 
\geq0
\label{condition}
\end{eqnarray}
where these coefficients are derived in (\ref{tensorcon}), (\ref{KV}), (\ref{CV}), (\ref{KS}), and (\ref{CS}). 

{\bf Solution I in case B5 }
	Let us now see the stability conditions for a special choice of model parameters. 
	To mimic the $\Lambda$CDM model at the background, we demand $M_{\rm eff}^2= 1$ in Eq.~(\ref{Meff}), which implies :
		\begin{eqnarray}
		G_2(Y) = g_2 \, (- Y)^{(2n + 1)}\,, 
		\quad G_3(Y) = g_3 \, (-Y)^{n}\,, \label{AFun}
		\end{eqnarray}
		with
		\begin{eqnarray}
		g_2 = \frac{3 g_3^2 n^2 }{2(1+2n)^2}  \,. 
		\end{eqnarray}
		In addition, we, for simplicity, choose $\alpha_7$ and $\alpha_8$ in order for
		making the sound speed of the vector and scalar modes irrelevant 
		to $A_*$, 
		\begin{eqnarray}
		&&
		\alpha_7 = -{a_7 \over Y^2}\,, \quad 
		\alpha_8 = \frac{a_8}{Y^2}\,.
		\end{eqnarray}
Then, we obtain the following conditions from  (\ref{condition})  :
\begin{eqnarray}
{\cal K}_V &=&{a_8{(16-a_8)} \over 32  A_*^{2}}>0\,, \\ 
{\cal K}_S &=&{9n^4{g_3^4(16-a_8)}  A_*^{2(1+4n)} \over {8} a_8 (1+2n)^2}>0\,,\\
c_V^2  &=&{8a_7\over a_8{(16-a_8)}}\geq0\,, \\
c_S^2  &=& {a_8-8 \over 3{(16-a_8)}(1+2n)}\geq0 \,, 
\label{model}
\end{eqnarray}
with ${\cal K}_T = 1/4$ and $c_T^2=1$.
	These conditions are translated into
\begin{eqnarray}
&0 < a_8 < 8  \qquad&\textrm{~~for~~}n<-1/2 \,,\label{para1}\\
&8  <  a_8<16  \qquad&\textrm{~~for~~}n>-1/2 \,.
\label{para2}
\end{eqnarray}
Here, $a_7>0$ and $g_3 \neq 0$ should be always satisfied. 
Thus, all perturbations are free of ghost and gradient instabilities 
for the parameters satisfying (\ref{para1}) and (\ref{para2}), 
and this is the first example of healthy cosmological solutions in degenerate theory of gravity.

{\bf Other cases }
First of all, as for solution II in the case $B5$, one can also check that all perturbations for tensor, vector, and scalar sectors given by (\ref{sol2.1}) and (\ref{sol2.2}) are also free of ghost 
and gradient instabilities with an appropriate choice of the arbitrary functions. 

Next, as discussed in \cite{Kimura:2016rzw}, the case $A4$ can be mapped to the (beyond) generalized Proca theories by using conformal and disformal metric transformations and vector field redefinition when matter fields are absent.
The cosmology and its perturbation analysis of the (beyond) generalized Proca
have already been investigated in \cite{2016JCAP...06..048D,2016arXiv160505066D,Heisenberg:2016eld,Nakamura:2017dnf}
and shown that all dynamical modes are free of ghost and gradient instabilities if arbitrary functions are appropriately chosen. 
We believe that stabilities do not change at least as long as the transformations of metric and vector field are invertible. Hence we conclude that the case $A4$ can be also a healthy branch.

As for other cases,  as long as tensor perturbations are concerned,
one will obtain formally the same result as (\ref{tensorcon}) irrespective of branches. This is because degenerate conditions do not change the structure of tensor perturbations at least around this cosmological background. However, in the cases $A1$, $A2$, $B2$, and $B3$ where $\alpha_1 = 1/Y$, tensor modes can be strongly coupled because ${\cal K}_T$ vanishes for this choice of $\alpha_1$ though the physical degrees of freedom in the tensor sector (as well as the vector sector) remain in a general background  \cite{Kimura:2016rzw}.
	Next, regarding vector perturbations, the kinetic term of $\widetilde{\delta A}_{I}$ vanishes in the cases $A1$, $A2$, $A3$, $B1$, $B2$, $B3$ and $B4$, and hence vector modes are strongly coupled around a cosmological background\footnote{
Strictly speaking, one cannot directly plug the degeneracy conditions
into (\ref{SV}) and (\ref{CV}) in the case $A1$, $A2$, $B2$ and $B3$. 
This is because the constraint equation of $S_I$ immediately gives $\delta A_I = 0$ in these cases. }.
In the Table.~\ref{table1}, we summarize vanishing or non-vanishing of the kinetic terms of the vector 
and tensor modes for each case.
\begin{table}[t]
	\centering
	\begin{tabular}{lccc}
		\hline
		Case  \quad  & \quad Vector \quad & \quad Tensor  \quad \\
		\hline \hline
		A1    & $\times$ & $\times$  \\
		A2 & $\times$ & $\times$  \\
		A3  & $\times$ & $\checkmark$  \\
		A4    &  $\checkmark$ & $\checkmark$ \\
		\hline 
		B1   & $\times$ & $\checkmark$   \\
		B2   & $\times$ & $\times$   \\
		B3  & $\times$ & $\times$ \\
		B4   &  $\times$ & $\checkmark$  \\	
		B5  & $\checkmark$ & $\checkmark$ \\
		B6     & $\checkmark$ & $\checkmark$ \\
		\hline
	\end{tabular}
	\caption{Summary of the tensor and vector perturbations for each model. ``$\checkmark$" and ``$\times$" respectively represent non-vanishing and vanishing of the kinetic term of the vector or tensor modes. 
	}
	\label{table1}
\end{table}
As one can see from the Table \ref{table1}, the case $B6$ remains as a possible candidate. However, a straightforward analysis shows that 
the kinetic term of the scalar perturbations vanishes 
after integrating out the lapse and shift variables in the spacially flat gauge.  
To understand whether this is a signal of strong coupling or partially massless case, one need to carefully investigate a constraint structure in a general background, 
and we will leave it for a future work.

\section{Conclusion}
We have studied cosmology of extended vector-tensor theories 
which were recently formulated in \cite{Kimura:2016rzw}, 
particularly focusing on the nature of cosmological perturbations 
around a homogeneous and isotropic background.
By deriving the second order action for linear perturbations, we found the vanishing of kinetic term among tensor, vector, and scalar sectors in theories except for the case A4 and B5, which implies a failure of linear perturbation theory around this cosmological background.
As investigated in \cite{Kimura:2016rzw}, 
the theory A4 can be mapped into (beyond) Proca theory by conformal 
and disformal metric transformations and vector field redefinition.
Provided that stabilities of perturbations do not change under such metric transformations, 
all perturbations in this case will be also free of ghost and gradient instabilities according to \cite{2016JCAP...06..048D,2016arXiv160505066D,Heisenberg:2016eld,Nakamura:2017dnf} as long as arbitrary functions are appropriately chosen.

As for the remaining non-trivial case B5, 
there are two branches in the background solutions, solution I and II respectively. 
Interestingly the solution I can be identical to $\Lambda$CDM model at least
 at the level of background equations under an appropriate choice of the arbitrary functions. 
Moreover the propagation speed of gravitational waves coincides with that of light
 when $\alpha_1=0$, which can be realized if conditions, $\alpha_2 = - 2/(3Y)$ and $\alpha_3 = 4/(3Y^2)$, are satisfied.
Then we have derived stability conditions for scalar perturbations to 
avoid ghost and gradient instabilities only in the presence of cosmological constant but in the absence of standard matter for simplicity. 
We have constructed a concrete model in which there is 
a viable parameter space satisfying all the stability conditions. 
Thus, we conclude that there exist healthy cosmological solutions in the non-trivial degenerate case B5 where all types of perturbations, namely scalar, vector and tensor perturbations, 
are free of ghost and gradient instabilities.
In addition, one can find self-accelerating solutions driven by $A_*$ in the solution I without ghost or gradient instabilities though it is not shown in this paper. 
This indicates that degenerate vector-tensor theory can be considered as a theoretically consistent candidate of mysterious dark energy.
In passing, we have also confirmed that solution II does not suffer from any instability in perturbations under an appropriate choice of arbitrary functions.

The result obtained above is in sharp contrast to the case of degenerate scalar-tensor theories, dubbed as DHOST \cite{2016JCAP...02..034L} or extended scalar-tensor theories \cite{2016JCAP...04..044C}. 
In these theories, either ghost or gradient instabilities in the scalar or tensor sector
is inevitable around a cosmological background, depending on the choice of free functions of the theory \cite{2016arXiv160408638D},
except for the cases that can be mapped from Horndeski theories via conformal and disformal transformations.
It should be noted that a crucial difference of the result in \cite{2016arXiv160408638D} and ours comes from the fact that degenerate vector-tensor theories are much broader than those scalar-tensor theories while some cases of vector-tensor theories reduce to degenerate scalar-tensor theories under a direct replacement such as $A_\mu \to \nabla_\mu \phi$. 
Our results indicate that such inevitable instabilities on a cosmological background is 
simply due to the specific structure of degenerate scalar-tensor theories themselves, and hence this does not imply a failure of a broad class of degenerate theories. 

\acknowledgments
R.~Kimura. and D.Y. thank Masahide Yamaguchi for many useful comments and discussions. 
R.~Kase and A.N. would like to thank the Yukawa Institute for Theoretical Physics at Kyoto University, where discussions during the YITP symposium YKIS2018a ``General Relativity -- The Next Generation --" were useful to complete this work.
This work was supported in part by JSPS Grant-in-Aid for Scientific Research 
Nos.~JP17K14297 (R.~Kase.), ~JP17K14276 (R.~Kimura.), JP25287054 (R.~Kimura.) and JSPS Postdoctoral Fellowships for Research Abroad (D.Y.). 
A.N. is supported in part by a JST grant ``Establishing a Consortium for the Development of Human Resources in Science and Technology''. The work of A.N. is also partly supported by the JSPS Grant-in-Aid for Scientific Research No.16H01092.

\bibliography{references}

\begin{thebibliography}{43}%
\makeatletter
\providecommand \@ifxundefined [1]{%
 \@ifx{#1\undefined}
}%
\providecommand \@ifnum [1]{%
 \ifnum #1\expandafter \@firstoftwo
 \else \expandafter \@secondoftwo
 \fi
}%
\providecommand \@ifx [1]{%
 \ifx #1\expandafter \@firstoftwo
 \else \expandafter \@secondoftwo
 \fi
}%
\providecommand \natexlab [1]{#1}%
\providecommand \enquote  [1]{``#1''}%
\providecommand \bibnamefont  [1]{#1}%
\providecommand \bibfnamefont [1]{#1}%
\providecommand \citenamefont [1]{#1}%
\providecommand \href@noop [0]{\@secondoftwo}%
\providecommand \href [0]{\begingroup \@sanitize@url \@href}%
\providecommand \@href[1]{\@@startlink{#1}\@@href}%
\providecommand \@@href[1]{\endgroup#1\@@endlink}%
\providecommand \@sanitize@url [0]{\catcode `\\12\catcode `\$12\catcode
  `\&12\catcode `\#12\catcode `\^12\catcode `\_12\catcode `\%12\relax}%
\providecommand \@@startlink[1]{}%
\providecommand \@@endlink[0]{}%
\providecommand \url  [0]{\begingroup\@sanitize@url \@url }%
\providecommand \@url [1]{\endgroup\@href {#1}{\urlprefix }}%
\providecommand \urlprefix  [0]{URL }%
\providecommand \Eprint [0]{\href }%
\providecommand \doibase [0]{http://dx.doi.org/}%
\providecommand \selectlanguage [0]{\@gobble}%
\providecommand \bibinfo  [0]{\@secondoftwo}%
\providecommand \bibfield  [0]{\@secondoftwo}%
\providecommand \translation [1]{[#1]}%
\providecommand \BibitemOpen [0]{}%
\providecommand \bibitemStop [0]{}%
\providecommand \bibitemNoStop [0]{.\EOS\space}%
\providecommand \EOS [0]{\spacefactor3000\relax}%
\providecommand \BibitemShut  [1]{\csname bibitem#1\endcsname}%
\let\auto@bib@innerbib\@empty
\bibitem [{\citenamefont {Will}(2014)}]{Will:2014kxa}%
  \BibitemOpen
  \bibfield  {author} {\bibinfo {author} {\bibfnamefont {C.~M.}\ \bibnamefont
  {Will}},\ }\href {\doibase 10.12942/lrr-2014-4} {\bibfield  {journal}
  {\bibinfo  {journal} {Living Rev. Rel.}\ }\textbf {\bibinfo {volume} {17}},\
  \bibinfo {pages} {4} (\bibinfo {year} {2014})},\ \Eprint
  {http://arxiv.org/abs/1403.7377} {arXiv:1403.7377 [gr-qc]} \BibitemShut
  {NoStop}%
\bibitem [{\citenamefont {Brout}\ \emph {et~al.}(1978)\citenamefont {Brout},
  \citenamefont {Englert},\ and\ \citenamefont {Gunzig}}]{Brout:1977ix}%
  \BibitemOpen
  \bibfield  {author} {\bibinfo {author} {\bibfnamefont {R.}~\bibnamefont
  {Brout}}, \bibinfo {author} {\bibfnamefont {F.}~\bibnamefont {Englert}}, \
  and\ \bibinfo {author} {\bibfnamefont {E.}~\bibnamefont {Gunzig}},\ }\href
  {\doibase 10.1016/0003-4916(78)90176-8} {\bibfield  {journal} {\bibinfo
  {journal} {Annals Phys.}\ }\textbf {\bibinfo {volume} {115}},\ \bibinfo
  {pages} {78} (\bibinfo {year} {1978})}\BibitemShut {NoStop}%
\bibitem [{\citenamefont {Starobinsky}(1979)}]{Starobinsky:1979ty}%
  \BibitemOpen
  \bibfield  {author} {\bibinfo {author} {\bibfnamefont {A.~A.}\ \bibnamefont
  {Starobinsky}},\ }\href@noop {} {\bibfield  {journal} {\bibinfo  {journal}
  {JETP Lett.}\ }\textbf {\bibinfo {volume} {30}},\ \bibinfo {pages} {682}
  (\bibinfo {year} {1979})},\ \bibinfo {note} {[Pisma Zh. Eksp. Teor.
  Fiz.30,719(1979)]}\BibitemShut {NoStop}%
\bibitem [{\citenamefont {Sato}(1981)}]{Sato:1980yn}%
  \BibitemOpen
  \bibfield  {author} {\bibinfo {author} {\bibfnamefont {K.}~\bibnamefont
  {Sato}},\ }\href@noop {} {\bibfield  {journal} {\bibinfo  {journal} {Mon.
  Not. Roy. Astron. Soc.}\ }\textbf {\bibinfo {volume} {195}},\ \bibinfo
  {pages} {467} (\bibinfo {year} {1981})}\BibitemShut {NoStop}%
\bibitem [{\citenamefont {Guth}(1981)}]{Guth:1980zm}%
  \BibitemOpen
  \bibfield  {author} {\bibinfo {author} {\bibfnamefont {A.~H.}\ \bibnamefont
  {Guth}},\ }\href {\doibase 10.1103/PhysRevD.23.347} {\bibfield  {journal}
  {\bibinfo  {journal} {Phys. Rev.}\ }\textbf {\bibinfo {volume} {D23}},\
  \bibinfo {pages} {347} (\bibinfo {year} {1981})}\BibitemShut {NoStop}%
\bibitem [{\citenamefont {Linde}(1982)}]{Linde:1981mu}%
  \BibitemOpen
  \bibfield  {author} {\bibinfo {author} {\bibfnamefont {A.~D.}\ \bibnamefont
  {Linde}},\ }\href {\doibase 10.1016/0370-2693(82)91219-9} {\bibfield
  {journal} {\bibinfo  {journal} {Phys. Lett.}\ }\textbf {\bibinfo {volume}
  {108B}},\ \bibinfo {pages} {389} (\bibinfo {year} {1982})}\BibitemShut
  {NoStop}%
\bibitem [{\citenamefont {{A. G. Riess et al.}}(1998)}]{Riess:aa}%
  \BibitemOpen
  \bibfield  {author} {\bibinfo {author} {\bibnamefont {{A. G. Riess et
  al.}}},\ }\href {\doibase 10.1086/300499} {\bibfield  {journal} {\bibinfo
  {journal} {Astron. J}\ }\textbf {\bibinfo {volume} {116}},\ \bibinfo {pages}
  {1009} (\bibinfo {year} {1998})},\ \Eprint
  {http://arxiv.org/abs/astro-ph/9805201} {astro-ph/9805201} \BibitemShut
  {NoStop}%
\bibitem [{\citenamefont {{S. Perlmutter et al.}}(1999)}]{Perlmutter:aa}%
  \BibitemOpen
  \bibfield  {author} {\bibinfo {author} {\bibnamefont {{S. Perlmutter et
  al.}}},\ }\href {\doibase 10.1086/307221} {\bibfield  {journal} {\bibinfo
  {journal} {Astrophys. J}\ }\textbf {\bibinfo {volume} {517}},\ \bibinfo
  {pages} {565} (\bibinfo {year} {1999})},\ \Eprint
  {http://arxiv.org/abs/astro-ph/9812133} {astro-ph/9812133} \BibitemShut
  {NoStop}%
\bibitem [{\citenamefont {Brans}\ and\ \citenamefont
  {Dicke}(1961)}]{PhysRev.124.925}%
  \BibitemOpen
  \bibfield  {author} {\bibinfo {author} {\bibfnamefont {C.}~\bibnamefont
  {Brans}}\ and\ \bibinfo {author} {\bibfnamefont {R.~H.}\ \bibnamefont
  {Dicke}},\ }\href {\doibase 10.1103/PhysRev.124.925} {\bibfield  {journal}
  {\bibinfo  {journal} {Phys. Rev.}\ }\textbf {\bibinfo {volume} {124}},\
  \bibinfo {pages} {925} (\bibinfo {year} {1961})}\BibitemShut {NoStop}%
\bibitem [{\citenamefont {{Clifton}}\ \emph {et~al.}(2012)\citenamefont
  {{Clifton}}, \citenamefont {{Ferreira}}, \citenamefont {{Padilla}},\ and\
  \citenamefont {{Skordis}}}]{Clifton:2011aa}%
  \BibitemOpen
  \bibfield  {author} {\bibinfo {author} {\bibfnamefont {T.}~\bibnamefont
  {{Clifton}}}, \bibinfo {author} {\bibfnamefont {P.~G.}\ \bibnamefont
  {{Ferreira}}}, \bibinfo {author} {\bibfnamefont {A.}~\bibnamefont
  {{Padilla}}}, \ and\ \bibinfo {author} {\bibfnamefont {C.}~\bibnamefont
  {{Skordis}}},\ }\href {\doibase 10.1016/j.physrep.2012.01.001} {\bibfield
  {journal} {\bibinfo  {journal} {Phys. Rept.}\ }\textbf {\bibinfo {volume}
  {513}},\ \bibinfo {pages} {1} (\bibinfo {year} {2012})},\ \Eprint
  {http://arxiv.org/abs/1106.2476} {arXiv:1106.2476} \BibitemShut {NoStop}%
\bibitem [{\citenamefont {Tsujikawa}(2010)}]{Tsujikawa:2011aa}%
  \BibitemOpen
  \bibfield  {author} {\bibinfo {author} {\bibfnamefont {S.}~\bibnamefont
  {Tsujikawa}},\ }\href {\doibase 10.1007/978-3-642-10598-2_3} {\bibfield
  {journal} {\bibinfo  {journal} {Lect. Notes Phys.}\ }\textbf {\bibinfo
  {volume} {800}},\ \bibinfo {pages} {99} (\bibinfo {year} {2010})},\ \Eprint
  {http://arxiv.org/abs/1101.0191} {arXiv:1101.0191 [gr-qc]} \BibitemShut
  {NoStop}%
\bibitem [{\citenamefont {Koyama}(2016)}]{Koyama:2015vza}%
  \BibitemOpen
  \bibfield  {author} {\bibinfo {author} {\bibfnamefont {K.}~\bibnamefont
  {Koyama}},\ }\href {\doibase 10.1088/0034-4885/79/4/046902} {\bibfield
  {journal} {\bibinfo  {journal} {Rept. Prog. Phys.}\ }\textbf {\bibinfo
  {volume} {79}},\ \bibinfo {pages} {046902} (\bibinfo {year} {2016})},\
  \Eprint {http://arxiv.org/abs/1504.04623} {arXiv:1504.04623 [astro-ph.CO]}
  \BibitemShut {NoStop}%
\bibitem [{\citenamefont {Vainshtein}(1972)}]{Vainshtein:1972aa}%
  \BibitemOpen
  \bibfield  {author} {\bibinfo {author} {\bibfnamefont {A.~I.}\ \bibnamefont
  {Vainshtein}},\ }\href {\doibase 10.1016/0370-2693(72)90147-5} {\bibfield
  {journal} {\bibinfo  {journal} {Phys. Lett.}\ }\textbf {\bibinfo {volume}
  {39B}},\ \bibinfo {pages} {393} (\bibinfo {year} {1972})}\BibitemShut
  {NoStop}%
\bibitem [{\citenamefont {Babichev}\ and\ \citenamefont
  {Deffayet}(2013)}]{Babichev:2013usa}%
  \BibitemOpen
  \bibfield  {author} {\bibinfo {author} {\bibfnamefont {E.}~\bibnamefont
  {Babichev}}\ and\ \bibinfo {author} {\bibfnamefont {C.}~\bibnamefont
  {Deffayet}},\ }\href {\doibase 10.1088/0264-9381/30/18/184001} {\bibfield
  {journal} {\bibinfo  {journal} {Class. Quant. Grav.}\ }\textbf {\bibinfo
  {volume} {30}},\ \bibinfo {pages} {184001} (\bibinfo {year} {2013})},\
  \Eprint {http://arxiv.org/abs/1304.7240} {arXiv:1304.7240 [gr-qc]}
  \BibitemShut {NoStop}%
\bibitem [{\citenamefont {{Ostrogradsky}}(1850)}]{Ostrogradsky1850}%
  \BibitemOpen
  \bibfield  {author} {\bibinfo {author} {\bibfnamefont {M.~V.}\ \bibnamefont
  {{Ostrogradsky}}},\ }\href@noop {} {\bibfield  {journal} {\bibinfo  {journal}
  {Mem. Acad. St. Petersbourg VI}\ }\textbf {\bibinfo {volume} {4}},\ \bibinfo
  {pages} {385} (\bibinfo {year} {1850})}\BibitemShut {NoStop}%
\bibitem [{\citenamefont {{Deffayet}}\ \emph {et~al.}(2009)\citenamefont
  {{Deffayet}}, \citenamefont {{Esposito-Far{\`e}se}},\ and\ \citenamefont
  {{Vikman}}}]{2009PhRvD..79h4003D}%
  \BibitemOpen
  \bibfield  {author} {\bibinfo {author} {\bibfnamefont {C.}~\bibnamefont
  {{Deffayet}}}, \bibinfo {author} {\bibfnamefont {G.}~\bibnamefont
  {{Esposito-Far{\`e}se}}}, \ and\ \bibinfo {author} {\bibfnamefont
  {A.}~\bibnamefont {{Vikman}}},\ }\href {\doibase 10.1103/PhysRevD.79.084003}
  {\bibfield  {journal} {\bibinfo  {journal} {Phys. Rev. D}\ }\textbf {\bibinfo
  {volume} {79}},\ \bibinfo {eid} {084003} (\bibinfo {year} {2009})},\ \Eprint
  {http://arxiv.org/abs/0901.1314} {arXiv:0901.1314 [hep-th]} \BibitemShut
  {NoStop}%
\bibitem [{\citenamefont {{Nicolis}}\ \emph {et~al.}(2009)\citenamefont
  {{Nicolis}}, \citenamefont {{Rattazzi}},\ and\ \citenamefont
  {{Trincherini}}}]{Nicolis:2008aa}%
  \BibitemOpen
  \bibfield  {author} {\bibinfo {author} {\bibfnamefont {A.}~\bibnamefont
  {{Nicolis}}}, \bibinfo {author} {\bibfnamefont {R.}~\bibnamefont
  {{Rattazzi}}}, \ and\ \bibinfo {author} {\bibfnamefont {E.}~\bibnamefont
  {{Trincherini}}},\ }\href {\doibase 10.1103/PhysRevD.79.064036} {\bibfield
  {journal} {\bibinfo  {journal} {Phys. Rev. D}\ }\textbf {\bibinfo {volume}
  {79}},\ \bibinfo {eid} {064036} (\bibinfo {year} {2009})},\ \Eprint
  {http://arxiv.org/abs/0811.2197} {arXiv:0811.2197 [hep-th]} \BibitemShut
  {NoStop}%
\bibitem [{\citenamefont {{Horndeski}}(1974)}]{Horndeski:1974aa}%
  \BibitemOpen
  \bibfield  {author} {\bibinfo {author} {\bibfnamefont {G.~W.}\ \bibnamefont
  {{Horndeski}}},\ }\href {\doibase 10.1007/BF01807638} {\bibfield  {journal}
  {\bibinfo  {journal} {Int. J. Theor. Phys.}\ }\textbf {\bibinfo {volume}
  {10}},\ \bibinfo {pages} {363} (\bibinfo {year} {1974})}\BibitemShut
  {NoStop}%
\bibitem [{\citenamefont {{Deffayet}}\ \emph {et~al.}(2011)\citenamefont
  {{Deffayet}}, \citenamefont {{Gao}}, \citenamefont {{Steer}},\ and\
  \citenamefont {{Zahariade}}}]{Deffayet:2011aa}%
  \BibitemOpen
  \bibfield  {author} {\bibinfo {author} {\bibfnamefont {C.}~\bibnamefont
  {{Deffayet}}}, \bibinfo {author} {\bibfnamefont {X.}~\bibnamefont {{Gao}}},
  \bibinfo {author} {\bibfnamefont {D.~A.}\ \bibnamefont {{Steer}}}, \ and\
  \bibinfo {author} {\bibfnamefont {G.}~\bibnamefont {{Zahariade}}},\ }\href
  {\doibase 10.1103/PhysRevD.84.064039} {\bibfield  {journal} {\bibinfo
  {journal} {Phys. Rev. D}\ }\textbf {\bibinfo {volume} {84}},\ \bibinfo {eid}
  {064039} (\bibinfo {year} {2011})},\ \Eprint {http://arxiv.org/abs/1103.3260}
  {arXiv:1103.3260 [hep-th]} \BibitemShut {NoStop}%
\bibitem [{\citenamefont {Kobayashi}\ \emph {et~al.}(2011)\citenamefont
  {Kobayashi}, \citenamefont {Yamaguchi},\ and\ \citenamefont
  {Yokoyama}}]{2011PThPh.126..511K}%
  \BibitemOpen
  \bibfield  {author} {\bibinfo {author} {\bibfnamefont {T.}~\bibnamefont
  {Kobayashi}}, \bibinfo {author} {\bibfnamefont {M.}~\bibnamefont
  {Yamaguchi}}, \ and\ \bibinfo {author} {\bibfnamefont {J.}~\bibnamefont
  {Yokoyama}},\ }\href {\doibase 10.1143/PTP.126.511} {\bibfield  {journal}
  {\bibinfo  {journal} {Prog. Theor. Phys.}\ }\textbf {\bibinfo {volume}
  {126}},\ \bibinfo {pages} {511} (\bibinfo {year} {2011})},\ \Eprint
  {http://arxiv.org/abs/1105.5723} {arXiv:1105.5723 [hep-th]} \BibitemShut
  {NoStop}%
\bibitem [{\citenamefont {Deffayet}\ and\ \citenamefont
  {Steer}(2013)}]{Deffayet:2013lga}%
  \BibitemOpen
  \bibfield  {author} {\bibinfo {author} {\bibfnamefont {C.}~\bibnamefont
  {Deffayet}}\ and\ \bibinfo {author} {\bibfnamefont {D.~A.}\ \bibnamefont
  {Steer}},\ }\href {\doibase 10.1088/0264-9381/30/21/214006} {\bibfield
  {journal} {\bibinfo  {journal} {Class. Quant. Grav.}\ }\textbf {\bibinfo
  {volume} {30}},\ \bibinfo {pages} {214006} (\bibinfo {year} {2013})},\
  \Eprint {http://arxiv.org/abs/1307.2450} {arXiv:1307.2450 [hep-th]}
  \BibitemShut {NoStop}%
\bibitem [{\citenamefont {Motohashi}\ \emph {et~al.}(2016)\citenamefont
  {Motohashi}, \citenamefont {Noui}, \citenamefont {Suyama}, \citenamefont
  {Yamaguchi},\ and\ \citenamefont {Langlois}}]{Motohashi:2016ftl}%
  \BibitemOpen
  \bibfield  {author} {\bibinfo {author} {\bibfnamefont {H.}~\bibnamefont
  {Motohashi}}, \bibinfo {author} {\bibfnamefont {K.}~\bibnamefont {Noui}},
  \bibinfo {author} {\bibfnamefont {T.}~\bibnamefont {Suyama}}, \bibinfo
  {author} {\bibfnamefont {M.}~\bibnamefont {Yamaguchi}}, \ and\ \bibinfo
  {author} {\bibfnamefont {D.}~\bibnamefont {Langlois}},\ }\href {\doibase
  10.1088/1475-7516/2016/07/033} {\bibfield  {journal} {\bibinfo  {journal}
  {JCAP}\ }\textbf {\bibinfo {volume} {1607}},\ \bibinfo {pages} {033}
  (\bibinfo {year} {2016})},\ \Eprint {http://arxiv.org/abs/1603.09355}
  {arXiv:1603.09355 [hep-th]} \BibitemShut {NoStop}%
\bibitem [{\citenamefont {Motohashi}\ \emph {et~al.}(2017)\citenamefont
  {Motohashi}, \citenamefont {Suyama},\ and\ \citenamefont
  {Yamaguchi}}]{Motohashi:2017eya}%
  \BibitemOpen
  \bibfield  {author} {\bibinfo {author} {\bibfnamefont {H.}~\bibnamefont
  {Motohashi}}, \bibinfo {author} {\bibfnamefont {T.}~\bibnamefont {Suyama}}, \
  and\ \bibinfo {author} {\bibfnamefont {M.}~\bibnamefont {Yamaguchi}},\
  }\href@noop {} {\  (\bibinfo {year} {2017})},\ \Eprint
  {http://arxiv.org/abs/1711.08125} {arXiv:1711.08125 [hep-th]} \BibitemShut
  {NoStop}%
\bibitem [{\citenamefont {Kimura}\ \emph
  {et~al.}(2017{\natexlab{a}})\citenamefont {Kimura}, \citenamefont
  {Sakakihara},\ and\ \citenamefont {Yamaguchi}}]{Kimura:2017gcy}%
  \BibitemOpen
  \bibfield  {author} {\bibinfo {author} {\bibfnamefont {R.}~\bibnamefont
  {Kimura}}, \bibinfo {author} {\bibfnamefont {Y.}~\bibnamefont {Sakakihara}},
  \ and\ \bibinfo {author} {\bibfnamefont {M.}~\bibnamefont {Yamaguchi}},\
  }\href {\doibase 10.1103/PhysRevD.96.044015} {\bibfield  {journal} {\bibinfo
  {journal} {Phys. Rev.}\ }\textbf {\bibinfo {volume} {D96}},\ \bibinfo {pages}
  {044015} (\bibinfo {year} {2017}{\natexlab{a}})},\ \Eprint
  {http://arxiv.org/abs/1704.02717} {arXiv:1704.02717 [hep-th]} \BibitemShut
  {NoStop}%
\bibitem [{\citenamefont {{Gleyzes}}\ \emph {et~al.}(2015)\citenamefont
  {{Gleyzes}}, \citenamefont {{Langlois}}, \citenamefont {{Piazza}},\ and\
  \citenamefont {{Vernizzi}}}]{2015PhRvL.114u1101G}%
  \BibitemOpen
  \bibfield  {author} {\bibinfo {author} {\bibfnamefont {J.}~\bibnamefont
  {{Gleyzes}}}, \bibinfo {author} {\bibfnamefont {D.}~\bibnamefont
  {{Langlois}}}, \bibinfo {author} {\bibfnamefont {F.}~\bibnamefont
  {{Piazza}}}, \ and\ \bibinfo {author} {\bibfnamefont {F.}~\bibnamefont
  {{Vernizzi}}},\ }\href {\doibase 10.1103/PhysRevLett.114.211101} {\bibfield
  {journal} {\bibinfo  {journal} {Physical Review Letters}\ }\textbf {\bibinfo
  {volume} {114}},\ \bibinfo {eid} {211101} (\bibinfo {year} {2015})},\ \Eprint
  {http://arxiv.org/abs/1404.6495} {arXiv:1404.6495 [hep-th]} \BibitemShut
  {NoStop}%
\bibitem [{\citenamefont {{Langlois}}\ and\ \citenamefont
  {{Noui}}(2016)}]{2016JCAP...02..034L}%
  \BibitemOpen
  \bibfield  {author} {\bibinfo {author} {\bibfnamefont {D.}~\bibnamefont
  {{Langlois}}}\ and\ \bibinfo {author} {\bibfnamefont {K.}~\bibnamefont
  {{Noui}}},\ }\href {\doibase 10.1088/1475-7516/2016/02/034} {\bibfield
  {journal} {\bibinfo  {journal} {JCAP}\ }\textbf {\bibinfo {volume} {2}},\
  \bibinfo {eid} {034} (\bibinfo {year} {2016})},\ \Eprint
  {http://arxiv.org/abs/1510.06930} {arXiv:1510.06930 [gr-qc]} \BibitemShut
  {NoStop}%
\bibitem [{\citenamefont {Langlois}\ and\ \citenamefont
  {Noui}(2016)}]{2015arXiv151206820L}%
  \BibitemOpen
  \bibfield  {author} {\bibinfo {author} {\bibfnamefont {D.}~\bibnamefont
  {Langlois}}\ and\ \bibinfo {author} {\bibfnamefont {K.}~\bibnamefont
  {Noui}},\ }\href {\doibase 10.1088/1475-7516/2016/07/016} {\bibfield
  {journal} {\bibinfo  {journal} {JCAP}\ }\textbf {\bibinfo {volume} {1607}},\
  \bibinfo {pages} {016} (\bibinfo {year} {2016})},\ \Eprint
  {http://arxiv.org/abs/1512.06820} {arXiv:1512.06820 [gr-qc]} \BibitemShut
  {NoStop}%
\bibitem [{\citenamefont {Ben~Achour}\ \emph
  {et~al.}(2016{\natexlab{a}})\citenamefont {Ben~Achour}, \citenamefont
  {Langlois},\ and\ \citenamefont {Noui}}]{2016arXiv160208398B}%
  \BibitemOpen
  \bibfield  {author} {\bibinfo {author} {\bibfnamefont {J.}~\bibnamefont
  {Ben~Achour}}, \bibinfo {author} {\bibfnamefont {D.}~\bibnamefont
  {Langlois}}, \ and\ \bibinfo {author} {\bibfnamefont {K.}~\bibnamefont
  {Noui}},\ }\href {\doibase 10.1103/PhysRevD.93.124005} {\bibfield  {journal}
  {\bibinfo  {journal} {Phys. Rev.}\ }\textbf {\bibinfo {volume} {D93}},\
  \bibinfo {pages} {124005} (\bibinfo {year} {2016}{\natexlab{a}})},\ \Eprint
  {http://arxiv.org/abs/1602.08398} {arXiv:1602.08398 [gr-qc]} \BibitemShut
  {NoStop}%
\bibitem [{\citenamefont {{Crisostomi}}\ \emph {et~al.}(2016)\citenamefont
  {{Crisostomi}}, \citenamefont {{Koyama}},\ and\ \citenamefont
  {{Tasinato}}}]{2016JCAP...04..044C}%
  \BibitemOpen
  \bibfield  {author} {\bibinfo {author} {\bibfnamefont {M.}~\bibnamefont
  {{Crisostomi}}}, \bibinfo {author} {\bibfnamefont {K.}~\bibnamefont
  {{Koyama}}}, \ and\ \bibinfo {author} {\bibfnamefont {G.}~\bibnamefont
  {{Tasinato}}},\ }\href {\doibase 10.1088/1475-7516/2016/04/044} {\bibfield
  {journal} {\bibinfo  {journal} {JCAP}\ }\textbf {\bibinfo {volume} {4}},\
  \bibinfo {eid} {044} (\bibinfo {year} {2016})},\ \Eprint
  {http://arxiv.org/abs/1602.03119} {arXiv:1602.03119 [hep-th]} \BibitemShut
  {NoStop}%
\bibitem [{\citenamefont {Ben~Achour}\ \emph
  {et~al.}(2016{\natexlab{b}})\citenamefont {Ben~Achour}, \citenamefont
  {Crisostomi}, \citenamefont {Koyama}, \citenamefont {Langlois}, \citenamefont
  {Noui},\ and\ \citenamefont {Tasinato}}]{BenAchour:2016fzp}%
  \BibitemOpen
  \bibfield  {author} {\bibinfo {author} {\bibfnamefont {J.}~\bibnamefont
  {Ben~Achour}}, \bibinfo {author} {\bibfnamefont {M.}~\bibnamefont
  {Crisostomi}}, \bibinfo {author} {\bibfnamefont {K.}~\bibnamefont {Koyama}},
  \bibinfo {author} {\bibfnamefont {D.}~\bibnamefont {Langlois}}, \bibinfo
  {author} {\bibfnamefont {K.}~\bibnamefont {Noui}}, \ and\ \bibinfo {author}
  {\bibfnamefont {G.}~\bibnamefont {Tasinato}},\ }\href {\doibase
  10.1007/JHEP12(2016)100} {\bibfield  {journal} {\bibinfo  {journal} {JHEP}\
  }\textbf {\bibinfo {volume} {12}},\ \bibinfo {pages} {100} (\bibinfo {year}
  {2016}{\natexlab{b}})},\ \Eprint {http://arxiv.org/abs/1608.08135}
  {arXiv:1608.08135 [hep-th]} \BibitemShut {NoStop}%
\bibitem [{\citenamefont {de~Rham}\ and\ \citenamefont
  {Matas}(2016)}]{2016arXiv160408638D}%
  \BibitemOpen
  \bibfield  {author} {\bibinfo {author} {\bibfnamefont {C.}~\bibnamefont
  {de~Rham}}\ and\ \bibinfo {author} {\bibfnamefont {A.}~\bibnamefont
  {Matas}},\ }\href {\doibase 10.1088/1475-7516/2016/06/041} {\bibfield
  {journal} {\bibinfo  {journal} {JCAP}\ }\textbf {\bibinfo {volume} {1606}},\
  \bibinfo {pages} {041} (\bibinfo {year} {2016})},\ \Eprint
  {http://arxiv.org/abs/1604.08638} {arXiv:1604.08638 [hep-th]} \BibitemShut
  {NoStop}%
\bibitem [{\citenamefont {Kimura}\ \emph
  {et~al.}(2017{\natexlab{b}})\citenamefont {Kimura}, \citenamefont {Naruko},\
  and\ \citenamefont {Yoshida}}]{Kimura:2016rzw}%
  \BibitemOpen
  \bibfield  {author} {\bibinfo {author} {\bibfnamefont {R.}~\bibnamefont
  {Kimura}}, \bibinfo {author} {\bibfnamefont {A.}~\bibnamefont {Naruko}}, \
  and\ \bibinfo {author} {\bibfnamefont {D.}~\bibnamefont {Yoshida}},\ }\href
  {\doibase 10.1088/1475-7516/2017/01/002} {\bibfield  {journal} {\bibinfo
  {journal} {JCAP}\ }\textbf {\bibinfo {volume} {1701}},\ \bibinfo {pages}
  {002} (\bibinfo {year} {2017}{\natexlab{b}})},\ \Eprint
  {http://arxiv.org/abs/1608.07066} {arXiv:1608.07066 [gr-qc]} \BibitemShut
  {NoStop}%
\bibitem [{\citenamefont {Heisenberg}(2014)}]{Heisenberg:2014rta}%
  \BibitemOpen
  \bibfield  {author} {\bibinfo {author} {\bibfnamefont {L.}~\bibnamefont
  {Heisenberg}},\ }\href {\doibase 10.1088/1475-7516/2014/05/015} {\bibfield
  {journal} {\bibinfo  {journal} {JCAP}\ }\textbf {\bibinfo {volume} {1405}},\
  \bibinfo {pages} {015} (\bibinfo {year} {2014})},\ \Eprint
  {http://arxiv.org/abs/1402.7026} {arXiv:1402.7026 [hep-th]} \BibitemShut
  {NoStop}%
\bibitem [{\citenamefont {Tasinato}(2014{\natexlab{a}})}]{Tasinato:2014eka}%
  \BibitemOpen
  \bibfield  {author} {\bibinfo {author} {\bibfnamefont {G.}~\bibnamefont
  {Tasinato}},\ }\href {\doibase 10.1007/JHEP04(2014)067} {\bibfield  {journal}
  {\bibinfo  {journal} {JHEP}\ }\textbf {\bibinfo {volume} {04}},\ \bibinfo
  {pages} {067} (\bibinfo {year} {2014}{\natexlab{a}})},\ \Eprint
  {http://arxiv.org/abs/1402.6450} {arXiv:1402.6450 [hep-th]} \BibitemShut
  {NoStop}%
\bibitem [{\citenamefont {Tasinato}(2014{\natexlab{b}})}]{Tasinato:2014mia}%
  \BibitemOpen
  \bibfield  {author} {\bibinfo {author} {\bibfnamefont {G.}~\bibnamefont
  {Tasinato}},\ }\href {\doibase 10.1088/0264-9381/31/22/225004} {\bibfield
  {journal} {\bibinfo  {journal} {Class. Quant. Grav.}\ }\textbf {\bibinfo
  {volume} {31}},\ \bibinfo {pages} {225004} (\bibinfo {year}
  {2014}{\natexlab{b}})},\ \Eprint {http://arxiv.org/abs/1404.4883}
  {arXiv:1404.4883 [hep-th]} \BibitemShut {NoStop}%
\bibitem [{\citenamefont {Beltran~Jimenez}\ and\ \citenamefont
  {Heisenberg}(2016)}]{Jimenez:2016isa}%
  \BibitemOpen
  \bibfield  {author} {\bibinfo {author} {\bibfnamefont {J.}~\bibnamefont
  {Beltran~Jimenez}}\ and\ \bibinfo {author} {\bibfnamefont {L.}~\bibnamefont
  {Heisenberg}},\ }\href {\doibase 10.1016/j.physletb.2016.04.017} {\bibfield
  {journal} {\bibinfo  {journal} {Phys. Lett.}\ }\textbf {\bibinfo {volume}
  {B757}},\ \bibinfo {pages} {405} (\bibinfo {year} {2016})},\ \Eprint
  {http://arxiv.org/abs/1602.03410} {arXiv:1602.03410 [hep-th]} \BibitemShut
  {NoStop}%
\bibitem [{\citenamefont {Heisenberg}\ \emph {et~al.}(2016)\citenamefont
  {Heisenberg}, \citenamefont {Kase},\ and\ \citenamefont
  {Tsujikawa}}]{Heisenberg:2016eld}%
  \BibitemOpen
  \bibfield  {author} {\bibinfo {author} {\bibfnamefont {L.}~\bibnamefont
  {Heisenberg}}, \bibinfo {author} {\bibfnamefont {R.}~\bibnamefont {Kase}}, \
  and\ \bibinfo {author} {\bibfnamefont {S.}~\bibnamefont {Tsujikawa}},\ }\href
  {\doibase 10.1016/j.physletb.2016.07.052} {\bibfield  {journal} {\bibinfo
  {journal} {Phys. Lett.}\ }\textbf {\bibinfo {volume} {B760}},\ \bibinfo
  {pages} {617} (\bibinfo {year} {2016})},\ \Eprint
  {http://arxiv.org/abs/1605.05565} {arXiv:1605.05565 [hep-th]} \BibitemShut
  {NoStop}%
\bibitem [{\citenamefont {Abbott}\ \emph
  {et~al.}(2017{\natexlab{a}})\citenamefont {Abbott} \emph
  {et~al.}}]{TheLIGOScientific:2017qsa}%
  \BibitemOpen
  \bibfield  {author} {\bibinfo {author} {\bibfnamefont {B.}~\bibnamefont
  {Abbott}} \emph {et~al.} (\bibinfo {collaboration} {Virgo, LIGO
  Scientific}),\ }\href {\doibase 10.1103/PhysRevLett.119.161101} {\bibfield
  {journal} {\bibinfo  {journal} {Phys. Rev. Lett.}\ }\textbf {\bibinfo
  {volume} {119}},\ \bibinfo {pages} {161101} (\bibinfo {year}
  {2017}{\natexlab{a}})},\ \Eprint {http://arxiv.org/abs/1710.05832}
  {arXiv:1710.05832 [gr-qc]} \BibitemShut {NoStop}%
\bibitem [{\citenamefont {Goldstein}\ \emph {et~al.}(2017)\citenamefont
  {Goldstein} \emph {et~al.}}]{Goldstein:2017mmi}%
  \BibitemOpen
  \bibfield  {author} {\bibinfo {author} {\bibfnamefont {A.}~\bibnamefont
  {Goldstein}} \emph {et~al.},\ }\href {\doibase 10.3847/2041-8213/aa8f41}
  {\bibfield  {journal} {\bibinfo  {journal} {Astrophys. J.}\ }\textbf
  {\bibinfo {volume} {848}},\ \bibinfo {pages} {L14} (\bibinfo {year}
  {2017})},\ \Eprint {http://arxiv.org/abs/1710.05446} {arXiv:1710.05446
  [astro-ph.HE]} \BibitemShut {NoStop}%
\bibitem [{\citenamefont {Abbott}\ \emph
  {et~al.}(2017{\natexlab{b}})\citenamefont {Abbott} \emph
  {et~al.}}]{Monitor:2017mdv}%
  \BibitemOpen
  \bibfield  {author} {\bibinfo {author} {\bibfnamefont {B.~P.}\ \bibnamefont
  {Abbott}} \emph {et~al.} (\bibinfo {collaboration} {Virgo, Fermi-GBM,
  INTEGRAL, LIGO Scientific}),\ }\href {\doibase 10.3847/2041-8213/aa920c}
  {\bibfield  {journal} {\bibinfo  {journal} {Astrophys. J.}\ }\textbf
  {\bibinfo {volume} {848}},\ \bibinfo {pages} {L13} (\bibinfo {year}
  {2017}{\natexlab{b}})},\ \Eprint {http://arxiv.org/abs/1710.05834}
  {arXiv:1710.05834 [astro-ph.HE]} \BibitemShut {NoStop}%
\bibitem [{\citenamefont {{De Felice}}\ \emph {et~al.}(2016)\citenamefont {{De
  Felice}}, \citenamefont {{Heisenberg}}, \citenamefont {{Kase}}, \citenamefont
  {{Mukohyama}}, \citenamefont {{Tsujikawa}},\ and\ \citenamefont
  {{Zhang}}}]{2016JCAP...06..048D}%
  \BibitemOpen
  \bibfield  {author} {\bibinfo {author} {\bibfnamefont {A.}~\bibnamefont {{De
  Felice}}}, \bibinfo {author} {\bibfnamefont {L.}~\bibnamefont
  {{Heisenberg}}}, \bibinfo {author} {\bibfnamefont {R.}~\bibnamefont
  {{Kase}}}, \bibinfo {author} {\bibfnamefont {S.}~\bibnamefont {{Mukohyama}}},
  \bibinfo {author} {\bibfnamefont {S.}~\bibnamefont {{Tsujikawa}}}, \ and\
  \bibinfo {author} {\bibfnamefont {Y.-l.}\ \bibnamefont {{Zhang}}},\ }\href
  {\doibase 10.1088/1475-7516/2016/06/048} {\bibfield  {journal} {\bibinfo
  {journal} {JCAP}\ }\textbf {\bibinfo {volume} {06}},\ \bibinfo {eid} {048}
  (\bibinfo {year} {2016})},\ \Eprint {http://arxiv.org/abs/1603.05806}
  {arXiv:1603.05806 [gr-qc]} \BibitemShut {NoStop}%
\bibitem [{\citenamefont {De~Felice}\ \emph {et~al.}(2016)\citenamefont
  {De~Felice}, \citenamefont {Heisenberg}, \citenamefont {Kase}, \citenamefont
  {Mukohyama}, \citenamefont {Tsujikawa},\ and\ \citenamefont
  {Zhang}}]{2016arXiv160505066D}%
  \BibitemOpen
  \bibfield  {author} {\bibinfo {author} {\bibfnamefont {A.}~\bibnamefont
  {De~Felice}}, \bibinfo {author} {\bibfnamefont {L.}~\bibnamefont
  {Heisenberg}}, \bibinfo {author} {\bibfnamefont {R.}~\bibnamefont {Kase}},
  \bibinfo {author} {\bibfnamefont {S.}~\bibnamefont {Mukohyama}}, \bibinfo
  {author} {\bibfnamefont {S.}~\bibnamefont {Tsujikawa}}, \ and\ \bibinfo
  {author} {\bibfnamefont {Y.-l.}\ \bibnamefont {Zhang}},\ }\href {\doibase
  10.1103/PhysRevD.94.044024} {\bibfield  {journal} {\bibinfo  {journal} {Phys.
  Rev.}\ }\textbf {\bibinfo {volume} {D94}},\ \bibinfo {pages} {044024}
  (\bibinfo {year} {2016})},\ \Eprint {http://arxiv.org/abs/1605.05066}
  {arXiv:1605.05066 [gr-qc]} \BibitemShut {NoStop}%
\bibitem [{\citenamefont {Nakamura}\ \emph {et~al.}(2017)\citenamefont
  {Nakamura}, \citenamefont {Kase},\ and\ \citenamefont
  {Tsujikawa}}]{Nakamura:2017dnf}%
  \BibitemOpen
  \bibfield  {author} {\bibinfo {author} {\bibfnamefont {S.}~\bibnamefont
  {Nakamura}}, \bibinfo {author} {\bibfnamefont {R.}~\bibnamefont {Kase}}, \
  and\ \bibinfo {author} {\bibfnamefont {S.}~\bibnamefont {Tsujikawa}},\ }\href
  {\doibase 10.1103/PhysRevD.95.104001} {\bibfield  {journal} {\bibinfo
  {journal} {Phys. Rev.}\ }\textbf {\bibinfo {volume} {D95}},\ \bibinfo {pages}
  {104001} (\bibinfo {year} {2017})},\ \Eprint
  {http://arxiv.org/abs/1702.08610} {arXiv:1702.08610 [gr-qc]} \BibitemShut
  {NoStop}%
\end{thebibliography}%

\end{document}